\documentstyle{aipproc}
\input epsf.tex   

\def \etal         {{et~al. }}

\def \gap
 {\mathrel{\hbox{\raise0.3ex\hbox{$>$}\kern-0.8em\lower0.8ex\hbox{$\sim$}}}}

\def \kms          {\hbox{km$\,$s$^{-1}$}}

\def\approxlt{\lower.2em\hbox{$\buildrel < \over \sim$}}
\def\approxgt{\lower.2em\hbox{$\buildrel > \over \sim$}}
\def \lap
 {\mathrel{\hbox{\raise0.3ex\hbox{$<$}\kern-0.75em\lower0.8ex\hbox{$\sim$}}}}

\def \ls           {\hbox{L$_{\odot}$}}

\def \ms           {\hbox{M$_{\odot}$}}           

\def \date	   {\ifcase\month \message{zero} \or
		    January \or February \or March \or April \or May \or June 
                    \or July \or 
		    August \or September \or October \or November \or 
		    December \fi
		    \space\number\day, \number\year}

\begin{document}
\title{Pre-Starbursts in Luminous IR Galaxies ?}

\author{Y. Gao$^*$, R. Gruendl$^*$, K.Y. Lo$^*$, C.Y. Hwang$^+$,
\& S. Veilleux$^{\dagger}$}
\address{$^*$Lab. for Astronomical Imaging, Dept. of Astronomy, 
Univ. of Illinois \\
$^+$IAA, Academia Sinica, Taiwan, ROC \ \ \ 
$^{\dagger}$Dept. of Astronomy, Univ. of Maryland}

\maketitle

\begin{abstract}
We present first results of our on-going BIMA Key Project:
imaging the CO(1-0) emission from a sample of 
10 LIRGs that are at various merging stages, with special emphasis
on systems apparently in the early/intermediate stages of merging.
We present here CO images with $\sim 5''$ resolution. An important 
result is the recognition of a plausible {\it pre-starburst} phase in 
some early LIRG mergers (e.g., Arp~302 and NGC~6670). Our initial 
analysis suggests that a merger-induced starburst phase may not 
begin before the nuclear
separation between the merging galaxies reaches roughly 10 kpc.
The surface gas density seems to increase from
a few times 10$^2 \ms {\rm pc}^{-2}$ to  
$>10^3 \ms {\rm pc}^{-2}$ while the prominent CO extent systematically 
decreases as merging progresses.
\end{abstract}

\section*{Introduction}

Luminous IR galaxies (LIRGs, $L_{\rm IR} \approxgt 2\times10^{11}\ls$, 
$H_0=75 \kms$ Mpc$^{-1}$), 
emit most of their bolometric luminosity in the far-IR (up to 
$\approxgt \ 90\%~$), and are the dominant class of galaxies
in the local universe at these high luminosities \cite{soi87}.
Many LIRGs are interacting/merging galaxies \cite{san88,lee94,mur96}
rich in molecular gas \cite{san91,gao96,sol97}. 
It is not well understood whether starbursts produce most of the IR 
luminosity, how the starbursts are initiated and what role 
galaxy-galaxy interactions might play in triggering these starbursts.
 
A study of the molecular gas properties
at various phases of the merging process in LIRGs would help
identify the key physical processes involved.
Previous CO imaging studies have concentrated on relatively
advanced merger systems \cite{sco91} in which the 
interstellar medium (ISM) has already been highly 
disrupted by the interaction and starbursts. 
 In order to isolate the conditions in the ISM
{\it leading} to starbursts, we have started a program to study a 
sample of LIRGs
chosen to represent different phases of the interacting/merging 
process, using the newly expanded 
Berkeley-Illinois-Maryland Association (BIMA) millimeter-wave array 
\cite{wel96}.
The goal is to sample statistically the evolution of physical conditions
of the molecular material in LIRGs as compared with the properties of the
IR emission along the merger sequence.
 
\section*{Sample and Observations}

Our sample emphasizes LIRGs which appear to be in the early and
intermediate 
stages of merging with large nuclear separation (Table 1). 
These LIRGs are potentially the most molecular gas-rich systems
since the CO luminosity is found to
increase with increasing separation of the merging nuclei in a sample of 
$\sim$ 50 LIRG mergers \cite{gao96}. 
 
The BIMA array is ideally 
suited to study these early/intermediate LIRG mergers given its 
large primary beam and wide spectral bandwidth.
The observations presented here were all made with the 9-element BIMA 
array in the H/C configurations in 1996. 
 
\begin{table}
\caption{Luminous Infrared Galaxies in a Merger Sequence.}
\label{table1}
\begin{tabular}{lcddddccc}
Source & cz & 
\multicolumn{1}{c}{R$_{\rm Sep}$\tablenote{Projected separation between 
the two galaxy nuclei.}} & L$_{IR}$ & 
\multicolumn{1}{c}{M(H$_2$)\tablenote{M(H$_2$)=4.78 L$_{\rm CO}$, where L$_{\rm CO}$ is the single-dish CO luminosity in K\kms~pc$^2$.}} & 
Beam &
\multicolumn{1}{c}{CO\tablenote{CO morphology, u$\equiv$ unresolved peak; e$\equiv$ extended structures resovled by the beam.}} &
\multicolumn{1}{c}{$\Sigma_{\rm H_2}$\tablenote{Observed peak surface gas density uncorrected for inclination. Note Arp~302 and NGC~6670 have inclination angles larger than 75$^{\circ}$}} & 
\multicolumn{1}{c}{SFE\tablenote{Given by L$_{\rm IR}$/M(H$_2$)[\ls/\ms].
The far-IR luminosities are estimated by scaling the far-IR emission and
extent following that of the radio continuum emission. First number is a
global value while the second row shows each peak value.}}\\
 & \kms & kpc & 10$^{11}$\ls & 10$^{10}$\ms & $''$ \ \ kpc & & \ms pc$^{-2}$ & \\
\tableline
ARP302&10166& 25.8 & 4.1 & 8.0 & 6.0 \ \ 3.7 &u+e& 1450 & 5.0    \\
      &     &      &     &     &     &   &      & 6.0,2.0 \\
N6670 & 8684& 14.6 & 3.8 & 5.5 & 4.8 \ \ 2.7 &u+e&  850 & 6.9     \\
      &     &      &     &     &     &   &      & 8.7,6.9 \\
U2369 & 9475& 13.1 & 3.9 & 3.4 & 6.1 \ \ 3.6 &u+e& 1250 & 11.5 \\
      &     &      &     &     &     &   &      & 15.4 \\
ARP55 &11773& 10.7 & 4.7 & 5.8 & 4.4 \ \ 3.1 &u+e& 1320 & 8.1  \\
IZw107&12043&  4.8 & 7.2 & 3.4 & 3.9 \ \ 2.9 &u+e& 1000 & 21.2 \\
      &     &      &     &     &     &   &      & 30.  \\
N5256 & 8239&  4.7 & 3.1 & 2.7 & 4.6 \ \ 2.3 &u+e& 1260 & 11.5 \\
N6090 & 8830&  3.5 & 3.0 & 2.4 & 5.8 \ \ 3.1 & u &  930 & 12.5 \\
      &     &      &     &     &     &   &      & 17.2 \\
\end{tabular}
\end{table}
 
\section*{Results and Discussions}

Fig.~1 shows integrated CO intensity (contours)
overlayed on broad-band images in 4 LIRGs in an order of decreasing 
nuclei separation. 
Although LIRGs in our sample have very small ranges in L$_{\rm IR}$ and
L$_{\rm CO}$, the apparent differences in CO morphology and gas properties
are clearly seen along the merger sequence:
 
\begin{figure}[b!] 
\epsfbox{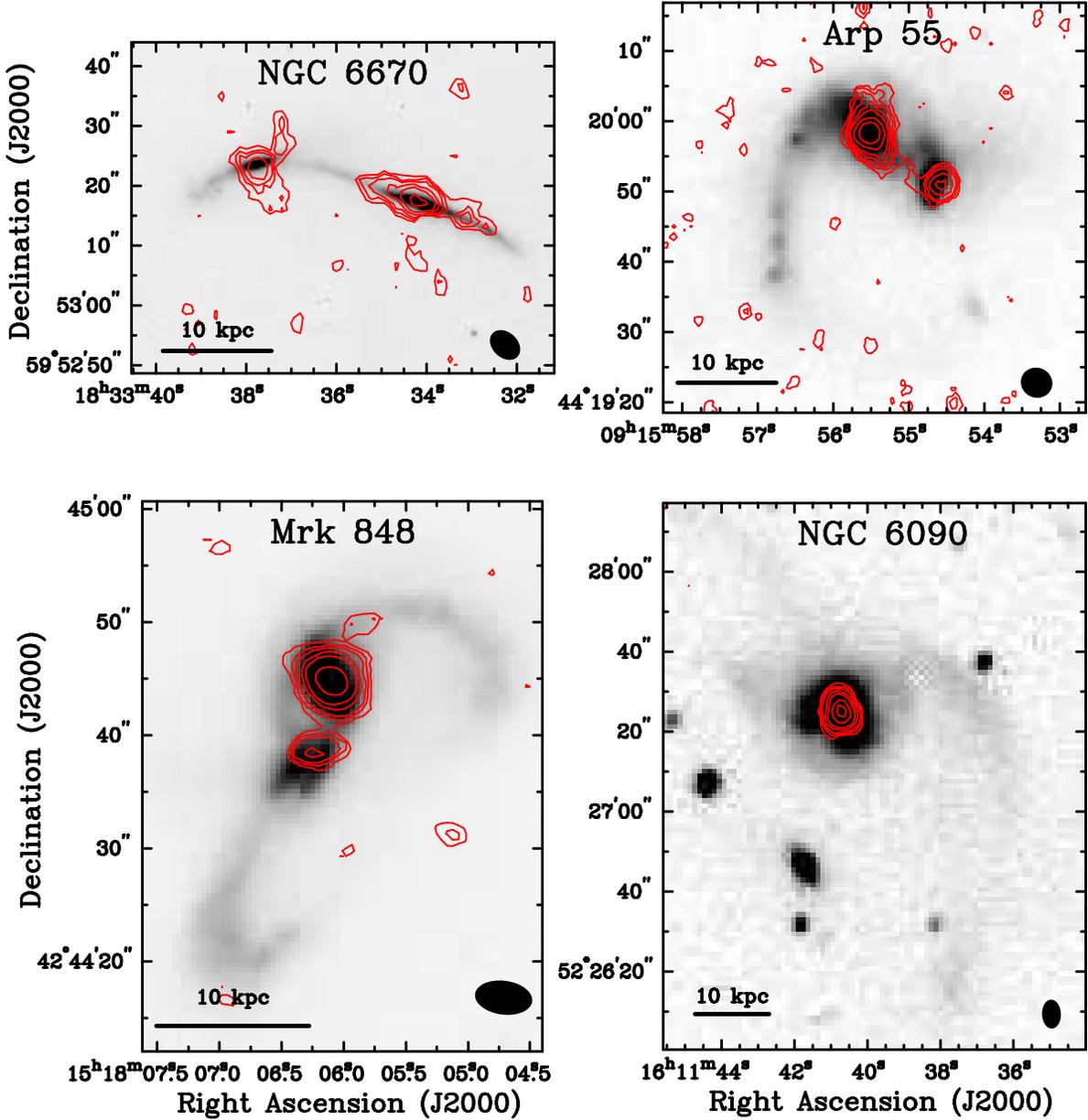}
\vspace{10pt}
\caption{CO contours overlayed on CCD images in four typical LIRGs 
in order of decreasing nuclear separation. The contours plotted 
are 2,3,4,6,8,12,16,24,32 $\sigma$ levels.}
\label{fig1}
\end{figure}

$\bullet$ The morphology of the molecular gas in LIRGs changes, 
along the
merger sequence, from the weakly disturbed two separated gas disks, 
e.g., early or pre-mergers like Arp~302 \cite{lo97} and NGC~6670
$\to$ the disturbed or merged-common-envelope gas disks 
(intermediate mergers like Arp~55 and Mrk~848) $\to$ a single  common
gas disk for the double nuclei of the two galaxies (close to the 
advanced mergers like NGC~6090).
 
$\bullet$
The total spatial CO extent drops from $\sim$ 20 kpc for the
early mergers to a few kpc for the intermediate and
advanced mergers. Very advanced mergers like Arp~220 have typical
nuclear CO concentration $\approxlt$ 1 kpc \cite{sco91}.
 
$\bullet$
The corrected face-on central gas surface density 
(lower-limits due to resolution) increases from a few times
10$^2$ \ms pc$^{-2}$ to $> 10^3$ \ms pc$^{-2}$ along the
sequence. Whereas advanced mergers such as Arp~220 and Mrk~273
have typical values $> 10^4$ \ms pc$^{-2}$.
 
$\bullet$
The ${\rm L_{IR}/M(H_2)}$ ratio (star formation efficiency, SFE) 
increases by roughly 
a factor of two from the early mergers to the intermediate/advanced 
mergers. When we scale the IR luminosity and extent with those of 
the radio continuum emission \cite{con90} using the 
well-known correlation between far-IR and radio continuum flux 
densities \cite{mar95}, we can estimate 
the central SFE ratio which tends to increase more drastically than 
the global SFE along the sequence (see Table~1).
 
$\bullet$ The SFE ratio usually ranges
from 20 to 100 $\ls/\ms$ in LIRGs. However, 
we found that early mergers like Arp~302 \cite{lo97} 
and NGC~6670 appear
to have much smaller SFE throughout the entire interacting/merging disks, 
comparable to that of GMCs in the Milky Way disk. 
The level of star formation activity in early mergers is therefore remarkably
similar to that of GMCs. This strongly 
suggests that LIRGs in the early stage of merging
are in a \underbar{pre-starburst} phase. 
  
$\bullet$ In intermediate stage LIRG mergers the starbursts 
appear to have 
``turned-on'', exhibited by higher SFE ratios especially in the
nuclei. This may imply that starburst phase does not begin before the 
separation between the merging galaxies reaches roughly 10 kpc  
characterizing these intermediate stage LIRG mergers.

\end{document}